\documentclass[twocolumn]{aastex631}
\usepackage{gensymb}
\usepackage{amsmath}	

\shorttitle{MSP in Terzan~6}
\shortauthors{Gao et al.}
\begin{document}

\title{Discovery of a millisecond pulsar associated with Terzan~6}

\correspondingauthor{Shi-Jie Gao \& Xiang-Dong Li}
\email{gaosj@smail.nju.edu.cn, lixd@nju.edu.cn}

\author[0000-0002-0822-0337]{Shi-Jie Gao}
\author[0000-0001-5684-0103]{Yi-Xuan Shao}
\affiliation{School of Astronomy and Space Science, Nanjing University, Nanjing, 210023, People's Republic of China}
\affiliation{Key Laboratory of Modern Astronomy and Astrophysics, Nanjing University, Ministry of Education, Nanjing, 210023, People's Republic of China}

\author[0000-0002-3386-7159]{Pei Wang}
\affiliation{National Astronomical Observatories, Chinese Academy of Sciences, Beijing, People's Republic of China}
\affiliation{Institute for Frontiers in Astronomy and Astrophysics, Beijing Normal University, Beijing 102206, People's Republic of China}

\author[0000-0002-5683-822X]{Ping Zhou}
\author[0000-0002-0584-8145]{Xiang-Dong Li}
\affiliation{School of Astronomy and Space Science, Nanjing University, Nanjing, 210023, People's Republic of China}
\affiliation{Key Laboratory of Modern Astronomy and Astrophysics, Nanjing University, Ministry of Education, Nanjing, 210023, People's Republic of China}

\author[0000-0001-8539-4237]{Lei Zhang}
\affiliation{National Astronomical Observatories, Chinese Academy of Sciences, Beijing, People's Republic of China}
\affiliation{CSIRO Astronomy and Space Science, P.O. Box 76, Epping, NSW 1710, Australia}

\author[0000-0002-3354-3859]{Joseph W. Kania}
\author[0000-0003-1301-966X]{Duncan R. Lorimer}
\affiliation{Center for Gravitational Waves and Cosmology, West Virginia University, Chestnut Ridge Research Building, Morgantown, WV, USA}
\affiliation{Department of Physics and Astronomy, West Virginia University, Morgantown, WV, USA}

\author[0000-0003-3010-7661]{Di Li}
\affiliation{Department of Astronomy, Tsinghua University, Beijing 100084, People's Republic of China}
\affiliation{National Astronomical Observatories, Chinese Academy of Sciences, Beijing, People's Republic of China}

\begin{abstract}
Observations show that globular clusters might be among the best places to find millisecond pulsars. However, the globular cluster Terzan 6 seems to be an exception without any pulsar discovered, although its high stellar encounter rate suggests that it harbors dozens of them. We report the discovery of the first radio pulsar, PSR~J1750--3116A, likely associated with Terzan~6 in a search of C-band (4--8~GHz) data from the Robert C. Byrd Green Bank Telescope with a spin period of 5.33~ms and dispersion measure, DM~$\simeq 383~{\rm pc\,cm^{-3}}$. The mean flux density of this pulsar is approximately 3~$\mu$Jy. The DM agrees well with predictions from the Galactic free electron density model, assuming a distance of $6.7~{\rm kpc}$ for Terzan~6. PSR~J1750--3116A is likely an isolated millisecond pulsar, potentially formed through dynamical interactions, considering the core-collapsed classification and the exceptionally high stellar encounter rate of Terzan~6. This is the highest radio frequency observation that has led to the discovery of a pulsar in a globular cluster to date. While L-band (1--2~GHz) observations of this cluster are unlikely to yield significant returns due to propagation effects, we predict that further pulsar discoveries in Terzan 6 will be made by existing radio telescopes at higher frequencies.
\end{abstract}


\section{Introduction}

Globular clusters (GCs) are gravitationally bound star clusters characterized by relatively high stellar densities toward their centers \citep{Gratton+2019}. The dense environment facilitates a high birth rate of low-mass X-ray binaries (LMXBs), which are the progenitors of millisecond pulsars \citep[MSPs; see, for example,][]{Ivanova+2013}. Consequently, GCs are ideal places for searching for pulsars compared to the Galactic disk. Currently, there are 330  pulsars known to be associated with GCs (see Paulo Friere’s catalog of Pulsars in Globular clusters\footnote{\url{https://www3.mpifr-bonn.mpg.de/staff/pfreire/GCpsr.html}\label{fn:gcpsr}}  for more details). While most of them are MSPs, which are expected given the age of GCs and their high specific incidence of LMXB, there are also several anomalous younger long-period ($\gtrsim 1~{\rm s}$) pulsars discovered in old GCs \citep[][]{1996ApJ...460L..41L,Boyles+2011,Zhou+2023,Wu+2023}.

\begin{figure}
\centering\includegraphics[width=\linewidth]{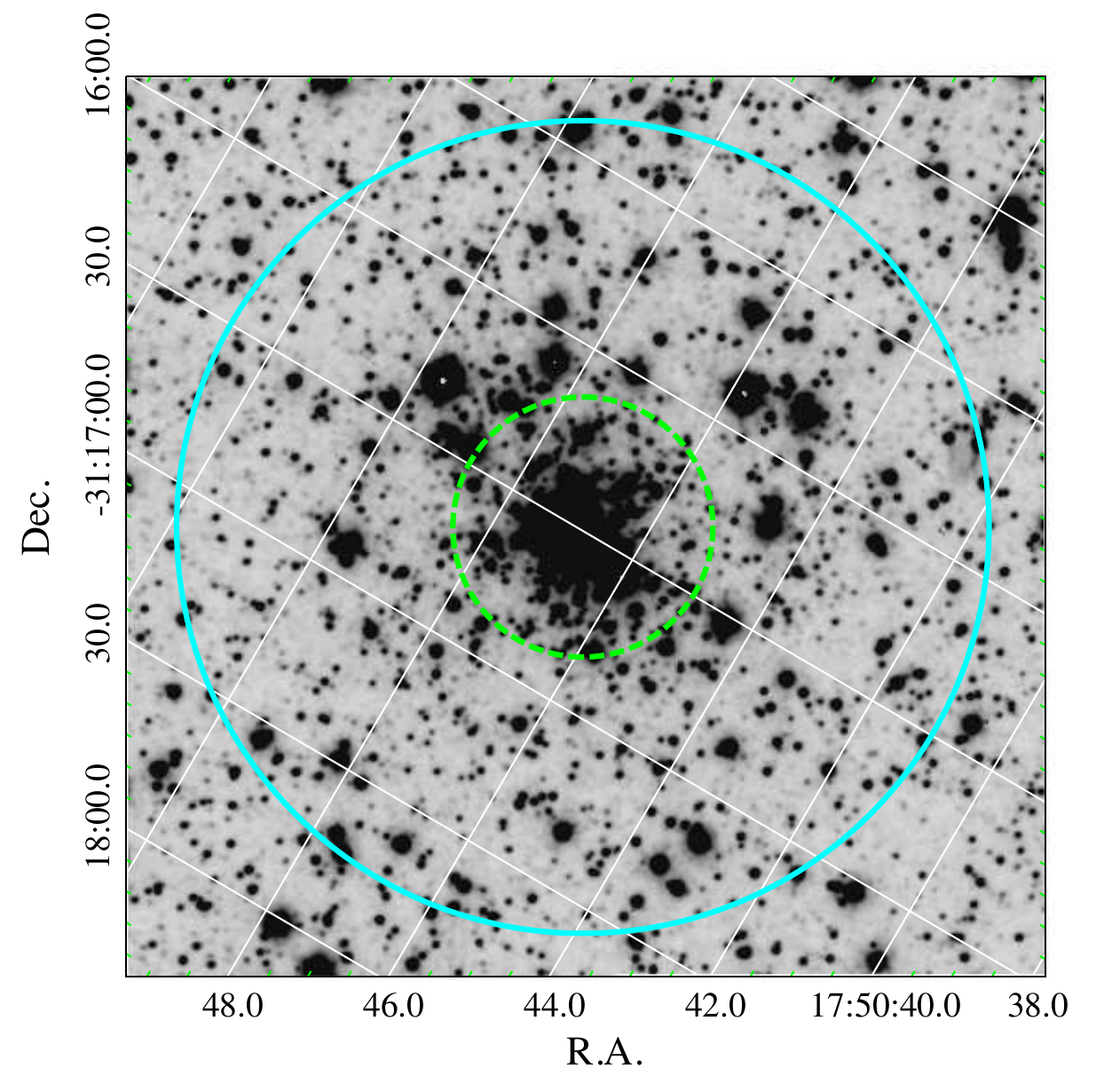}
    \caption{Ks-band image of Terzan~6 \citep{Minniti+VVVX}. The green circle represents the half-light radius ($0.4'$) of Terzan~6, and the cyan circle the beam size ($2.5'$, full-width at half-maximum) of the GBT's C-band receiver.\label{fig:terzan6}}
\end{figure}

Terzan~6 is a metal-rich core-collapsed Galactic GC \citep{Trager+1995,Barbuy+1997} centered at $\alpha_{\rm J2000}=17^{\rm h}50^{\rm m}46^{\rm s}.854$, $\delta_{\rm J2000}=-31{\degree}16'29''.384$ \citep{intZand+2003}. Optical and near-infrared photometric observations indicate the distance to Terzan~6 to be $6.7~{\rm kpc}$ \citep{Fahlman+1995,Barbuy+1997,Valenti+2007}. \autoref{fig:terzan6} shows the Ks-band image of Terzan~6 \citep{Minniti+VVVX}. The core and half-light radii (green circle in \autoref{fig:terzan6}) of Terzan~6 are $3''$ and $0'.4$, respectively \citep{Harris+1996}. Located in the direction of the Galactic bulge, Terzan~6 is $2.2\degree$ below the Galactic plane with a high extinction $E(B-V)=2.35$ \citep{Valenti+2007}.

\begin{table*}
    \setlength{\tabcolsep}{20pt}
    \caption{GCs with the highest $\Gamma$ values, providing their distances $d$ \citep{Harris+1996}, $\Gamma$ values \citep{Bahramian+2013}, the number $N_{\rm psr}$ of known pulsars, the number of known isolated pulsars $N_{\rm iso}$ and the expected number $\lambda$ of potential radio pulsars. \label{tab:GCs}}
    \begin{tabular}{llllll}
    \hline
    GCs $^a$&$d~{\rm (kpc)}$&$\Gamma$ $^b$&$N_{\rm psr}$ $^ c$&$N_{\rm iso}$ $^c$&$\lambda$ $^d$\\
    \hline
    Terzan~5&6.9&6800&49&20&104\\
    *NGC~7078 (M~15)&10.4&4510&14&13&80\\
    NGC~6715 (M~54)&26.5&2520&0&0&55\\
    *Terzan~6&6.7&2470&0&0&54\\
    NGC~6441&11.6&2300&9&6&52\\
    *NGC~6266 (M62)$^{e}$&6.8&1670&10&0&42\\
    NGC~1851&12.1&1530&15&6&40\\
    NGC~6440&8.5&1400&8&4&37\\
    *NGC~6624&7.9&1150&12&10&33\\
    *NGC~6681 (M~70)&9.0&1040&3&2&31\\
    NGC~104 (47~Tuc)&4.5&1000&36&12&30\\
    \hline
    \end{tabular}
    \tablecomments{$^a$~Core-collapsed GCs are indicated by``*" \citep{Harris+1996}.
    $^b$~$\Gamma$ values are normalized to give 47~Tuc's $\Gamma=1000$.
    $^c$~Numbers of known pulsars are referred to Paulo Friere’s catalog\textsuperscript{\ref{fn:gcpsr}}.
    $^d$ $\ln \lambda=-1.1+1.5\log \Gamma$ \citep{Turk+2013}.
    $^e$~NGC~6266 is either core-collapsed or on the verge of collapsing, see discussion in \cite{Vleeschower+2024}.}
\end{table*}

The stellar encounter rate $\Gamma\propto \rho_0^{1.5}r_{\rm c}$ (where $\rho_0$ is the central mass density and $r_{\rm c}$ is the core radius) is an important indicator for characterizing the dynamics of GCs. It is well known that  $\Gamma$ correlates with the number of pulsars within a GC \citep{Pooley+2003,Abdo+2009}, and the expected
number $\lambda$ of potential radio pulsars could be estimated from the following relation suggested by \cite{Turk+2013}: $\ln\lambda=-1.1+1.5\log \Gamma$. 
\autoref{tab:GCs} summarizes the measured and inferred parameters of some Galactic GCs with high $\Gamma$ values. Here the $\Gamma$ values are normalized such that {47~Tuc} has $\Gamma=1000$ \citep{Bahramian+2013}. Terzan~6 is among the GCs with highest $\Gamma$ values, placing it as a favorable host of LMXBs, MSPs, and {merger}/accretion-induced collapse products \citep[e.g.,][]{Verbunt+1987,Pooley+2003,Verbunt+2014,Kremer+2021,Kremer+2022,Kremer+2023,Ye+2024}. 
According to the $\lambda-\Gamma$ relation, Terzan~6 is expected to harbor around 54 radio pulsars. GCs with similar $\Gamma$ values with Terzan~6 such as Terzan~5 \citep{Padmanabh+2024}, M15 \citep{Wu+2023} and NGC~6441\footnote{\url{https://trapum.org/discoveries/}}, have 49, 14 and 9 confirmed pulsars, respectively. Although two X-ray sources have been indeed identified in Terzan~6, i.e., the transient LMXB GRS~1747--312 \citep{Predehl+1991} and the eclipsing LMXB burster Terzan~6~X2 \citep{vandenBerg+2024}, no MSPs have been detected so far. A similar situation occurs in NGC~6715, probably due to its very large distance of $26.5~{\rm kpc}$, while Terzan~6 is the second nearest GC listed in \autoref{tab:GCs}. 

Here we report the discovery a $5.33~{\rm ms}$ pulsar associated with Terzan~6, referred to as PSR~J1750--3116A. The rest of this letter is structured as follows. We describe the observations and data reduction processes in \autoref{sec:obs}.  The search results are presented in \autoref{sec:res}. We discuss the implications of these results in \autoref{sec:dis}. Our conclusions are provided in \autoref{sec:con}.

\begin{figure*}
    \centering
    \includegraphics[width=\linewidth]{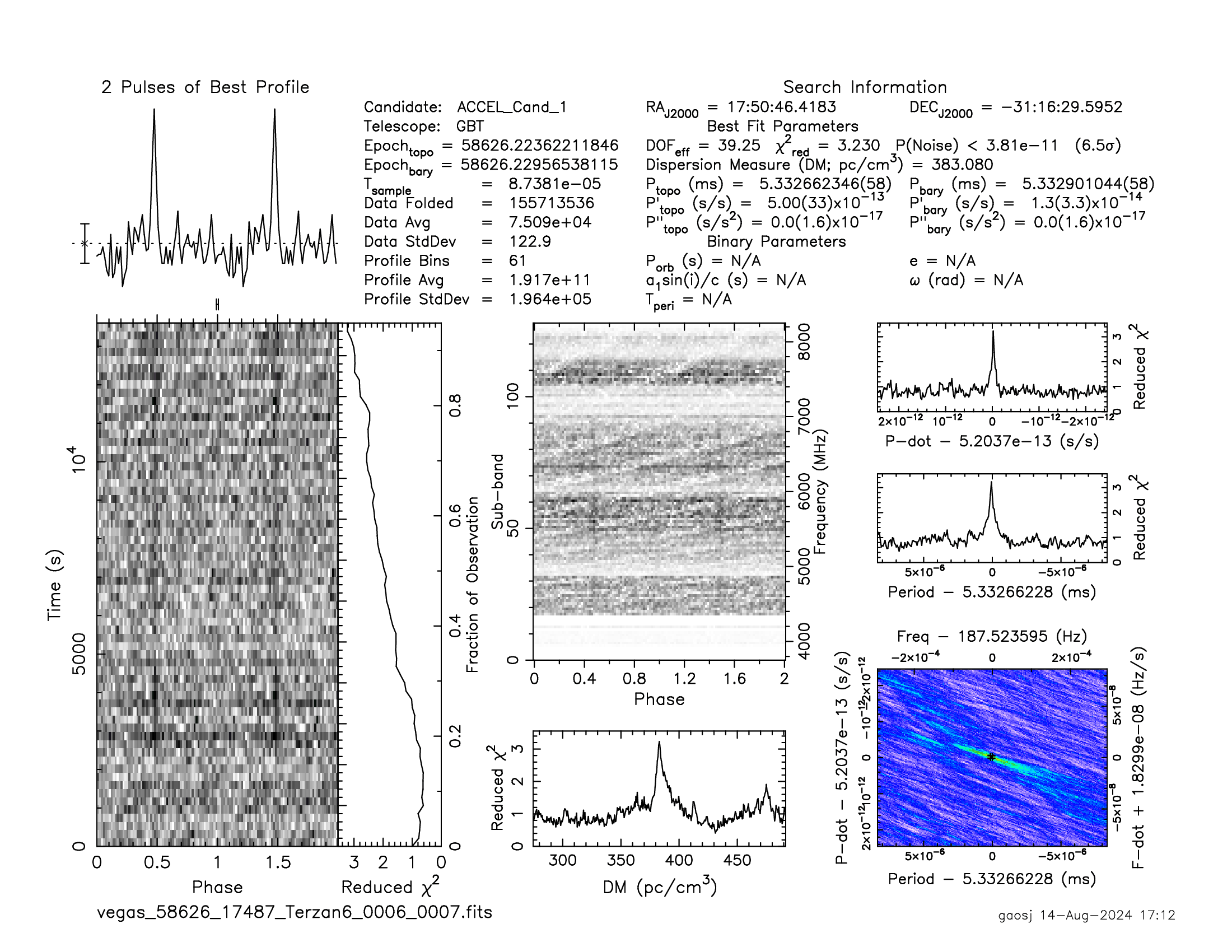}
    \caption{Discovery plot of PSR~J1750--3116A from the second observation on MJD~58626, generated by \texttt{PRESTO} routine \texttt{prepfold}. The lower-left and upper-middle panels show the phase vs. time and phase vs. observing frequency, respectively. The cumulative profile is plotted at the top of the left panel and the sidebar to the right of the left panel shows the increase in the reduced-$\chi^2$ (reflecting the increased level of statistical significance) with observing time. The lower-middle panel shows the reduced-$\chi^2$ vs. dispersion measure, {with} a peak located at DM~$=383.08~{\rm pc~cm^{-3}}$. The right panels shows the reduced-$\chi^2$ respect to trial period and period derivative, where the highest values of reduced-$\chi^2$ indicate the best-fit of observed period and period derivative.\label{fig:obs2}}
\end{figure*}

\begin{figure*}
    \centering
    \includegraphics[width=\linewidth]{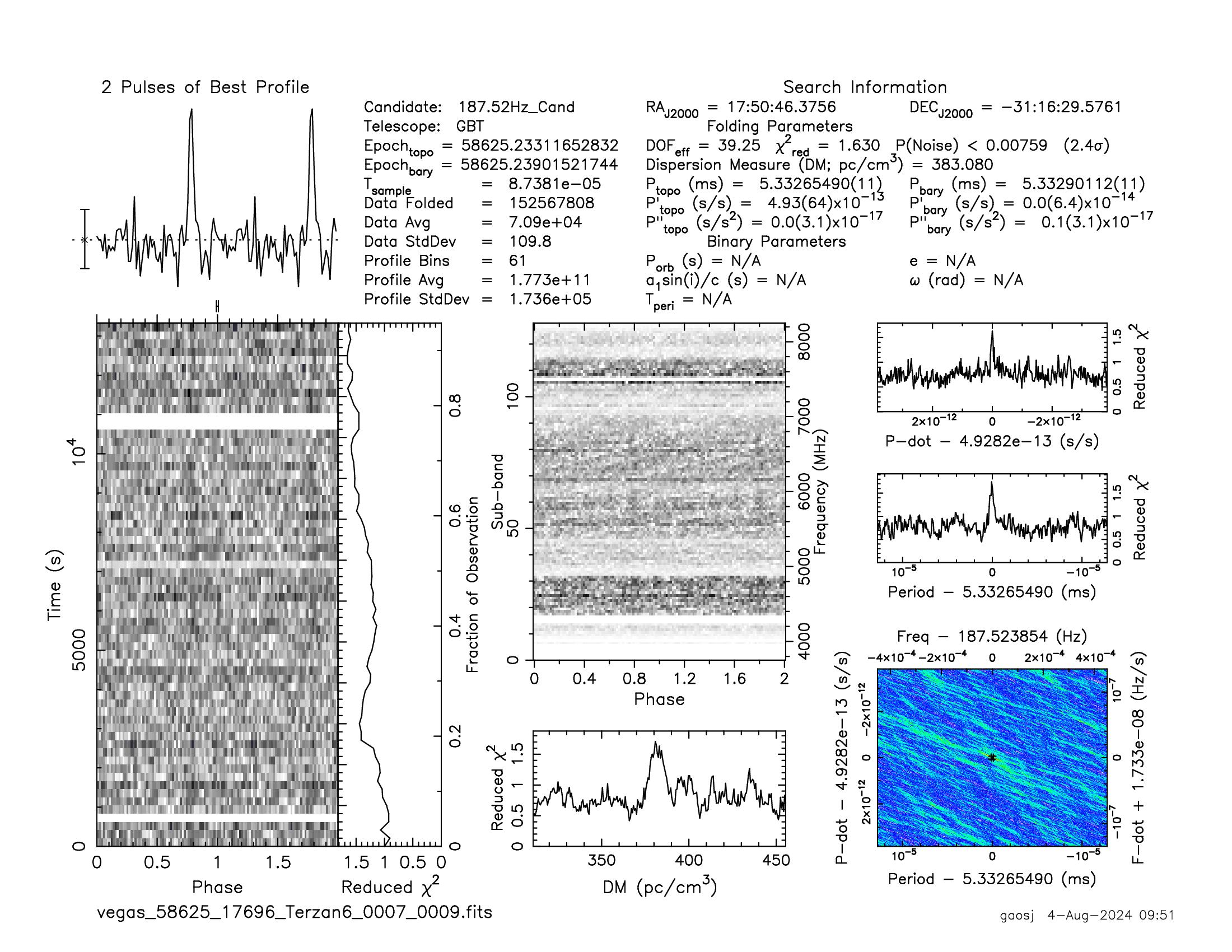}
    \caption{Similar to \autoref{fig:obs2} but for the first observation on MJD~58625.\label{fig:obs1}}
\end{figure*}

\section{Observations and Data reduction}\label{sec:obs}
Observations of Terzan~6 were conducted with the Robert C. Byrd Green Bank Telescope \citep[GBT;][]{Prestage+2009} using the C-band receiver on two occasions: MJD~58625 (UTC 2019-05-22) and MJD~58626 (UTC 2019-05-23) (project ID: AGBT19A\_477, PI: Duncan Lorimer). The beam size (full-width at half-maximum) of the C-band receiver is $2.5'$ (cyan circle in \autoref{fig:terzan6})\footnote{\url{https://www.gb.nrao.edu/scienceDocs/GBTpg.pdf}}, providing a good coverage of Terzan~6. Data were recorded with 3072 frequency channels covering a total effective bandwidth $3.8~{\rm GHz}$ centered at $6~{\rm GHz}$. The time resolution was $87.381~{\rm \mu s}$ and the integration time for both observations was $\sim4.6$ hours. The system temperature of the GBT's C-band receiver increased by several K at the beginning and end of $\sim0.5$ hour of the observations, likely due to atmospheric opacity when Terzan~6 (Decl. $\sim -31\degree$) was visible at low elevations. Consequently, the first and last $\sim 500-2000~{\rm s}$ observational data were excluded in the following analysis.

We searched for periodic dispersed pulsations using the PulsaR Exploration and Search TOolkit\footnote{\url{https://github.com/scottransom/presto}} \citep[\texttt{PRESTO},][]{Ransom+2011}. We first used the \texttt{PRESTO} routine \texttt{rfifind} to mask and zap radio-frequency interference (RFI) in both the time and frequency domains. Then the search-mode PSRFITS-format\footnote{\url{https://www.atnf.csiro.au/research/pulsar/psrfits_definition/Psrfits.html}} data were de-dispersed with the \texttt{PRESTO} routine \texttt{prepsubband}\footnote{Utilizing a GPU-version (\url{https://github.com/zdj649150499/Presto_GPU}) running on a NVIDIA$^\circledR$ A100 GPU.} to form barycentric time-series in a dispersion measure (DM) range of $0-1000~{\rm pc~cm^{-3}}$, with a step of $0.5~{\rm pc~cm^{-3}}$, following the \texttt{PRESTO} Python script \texttt{DDplan.py}. Subsequently, the \texttt{PRESTO} routine \texttt{accelsearch} was employed to search for periodic signals with a Fourier-domain acceleration search technique \citep{Ransom+2002}. The  summed number of harmonics was up to 16 and the maximum Fourier frequency derivative to search was set to $\texttt{zmax}=\dot f T_{\rm obs}^2=100$, where $\dot f$ and $T_{\rm obs}$ are the spin frequency derivative and the total observation length, respectively. This corresponds to a maximum drift of the physical linear acceleration $a=c\dot f/f=\texttt{zmax}\times c/\left(fT_{\rm obs}^2\right)$, where $c$ is the speed of light in vacuum and $f$ is the spin frequency. Candidates generated by \texttt{accelsearch} were winnowed using the Python script \texttt{ACCEL\_sift.py} of \texttt{PRESTO} with a sigma threshold greater than $4.0$. Finally, the folded diagnostic plots were produced by the \texttt{PRESTO} routine \texttt{prepfold} and were inspected by eye to identify the promising candidates.

\section{Results}\label{sec:res}

We found a faint pulsar candidate PSR~J1750--3116A with a spin period $P\simeq5.33~{\rm ms}$ and DM~$\simeq 383~{\rm  pc~cm^{-3}}$ in the second observation on MJD~58626. \autoref{fig:obs2} shows the folding results generated by the \texttt{PRESTO} routine \texttt{prepfold}. The significance of the detection is computed by calculating the reduced-$\chi^2$ statistic for a model assuming no pulsations \citep{Leahy+1983}. The reduced-$\chi^2$ is $3.230$ with an effective degrees of freedom of $39.25$ \citep[for details, see Appendix~E of][]{Bachetti+2021}. The probability that the signal might be due to noise is $<3.81\times 10^{-11}$, which is expressed in terms of equivalent Gaussian significance of 6.5-$\sigma$. The lower-left and upper-middle panels show the phase vs. time and phase vs. observing frequency relations, respectively, where the pulsar features (two vertical lines) can be easily identified. The cumulative profile is plotted on the top of the left panel. The sidebar at the right of the left panel shows the increase in the reduced-$\chi^2$ with the observing time. The lower-middle panel illustrates the dispersion measure vs. reduced-$\chi^2$, with a peak located at $\simeq 383.08~{\rm pc~cm^{-3}}$.
The right panels show the reduced-$\chi^2$ with respect to period and period derivative, where the highest values of reduced-$\chi^2$ indicate the best-fit of the observed period and period derivative.

We failed to detect the $5.33~{\rm ms}$ signal during a blind search of the first observation on MJD~58625. Using the folding results of the second observation, however, we successfully folded the data with the same DM and $P$ during the first observation, which is shown in \autoref{fig:obs1}. Although the signal in the first observation is much weaker than on MJD~58626, when folded using the parameters found above, the pulsar-like features are clearly discernible. The dimming in the first observation may be due to scintillation or intrinsic emission variations of the pulsar.

Neither significant acceleration (the \texttt{PRESTO} routine \texttt{accelsearch} gives $-0.00185(93)~{\rm m\,s^{-2}}$ for the first observation) nor barycentric period variation between the two observations was detected, suggesting that the pulsar is either isolated or in a long-period orbit.
{An isolated pulsar seems to be consistent with the Terzan~6's classification as a core-collapsed cluster, where dynamical interaction processes are more frequent and intense. Isolated MSPs could be formed from neutron star--main-sequence star tidal disruption events \citep{Kremer+2022,Ye+2024}, gravitational collapse of  double white dwarf merger remnants \citep{Ye+2024}, and disruption of MSP binaries by dynamical interactions during core collapse of the cluster \citep{Verbunt+2014}.}
As shown in \autoref{tab:GCs}, core-collapsed GCs generally have high ratios of $N_{\rm iso}/N_{\rm psr}$, e.g., $13/14$ for NGC~7078 (M15) and $10/12$ for NGC~6624. Whether PSR~J1750--3116A is isolated needs to be confirmed or falsified by further follow-up observations. Nevertheless, assuming that the pulsar is isolated, we tried to obtain the spin frequency through connecting the two observations. We used the \texttt{PRESTO} Python scripts \texttt{pygaussfit.py} and \texttt{get\_TOAs.py} to form a standard reference template profile and extracted the times of arrival (TOAs) of these two observations, respectively. We also fixed the location of PSR~J1750--3116A according to the pointing of the beam of the second observation and DM. Using \texttt{TEMPO2}\footnote{\url{https://sourceforge.net/projects/tempo2}} \citep{Nice+2015} the spin frequency was fitted  to be $187.515196274(72)~{\rm Hz}$. The input and output parameters are summarized in \autoref{tab:timing}.

\begin{table*}
    \setlength{\tabcolsep}{45pt}
    \caption{\texttt{TEMPO2}'s input and output parameters for PSR~J1750--3116A.\label{tab:timing}}
    \begin{tabular}{lll}
    \hline
    PSR&&J1750--3116A\\
    \hline
         R.A. (J2000, hh:mm:ss)&fixed&17:50:46(11.7) $^{\rm a}$ \\
         Decl. (J2000, dd:mm)&fixed&--31:16(2.5) $^{\rm a}$ \\
         Dispersion measure &fixed& $383.08~{\rm pc~cm^{-3}}$\\
         Spin frequency derivative&fixed&$0.0$ \\
        Spin frequency&fitted&$187.515196274(72)~{\rm Hz}$\\
        Spin period&derived&${0.0053329011188(20)}~{\rm s}$\\
        Timing epoch&$\cdots$& MJD~58626\\
        Total time span&$\cdots$&1.128~{\rm d}\\
        Number of TOAs&$\cdots$&13\\
        Post-fit residual&$\cdots$&$48.495~{\rm \mu s}$\\
    \hline
    \end{tabular}
    \tablecomments{$^{\rm a}$ The uncertainties for the position are set to the beam size ($2.5'$) of the GBT's C-band receiver. During fitting, the position is set to the center coordinates (R.A., Decl.) = ($17{\rm ^h}50{\rm ^m}46{\rm ^s}.42$, --31$\degree16'30''.00$), which is where the GBT's C-band receiver was pointed during the second observation.}
\end{table*}

\section{Discussion} \label{sec:dis}

The high DM of $383~{\rm pc~cm^{-3}}$ for PSR~J1750--3116A places Terzan~6 as the GC {with} the second highest DM to date behind GLIMPSE-C01~\citep[with DM~$\sim 450-520~{\rm pc~cm^{-3}}$;][]{McCarver+2024}. Adopting a distance of $6.7~{\rm kpc}$ to Terzan~6, the Galactic electron density models YMW2016 \citep{Yao+2017} and NE2001 \citep{Cordes+2002} predict the DM values to be $271~{\rm pc~cm^{-3}}$ and $397~{\rm pc~cm^{-3}}$, respectively\footnote{Calculated using \texttt{PyGEDM} \citep{pygedm} web app (\url{https://apps.datacentral.org.au/pygedm/}).}. The measured DM for PSR~J1750--3116A falls between them, strongly supporting the association between PSR~J1750--3116A and Terzan~6.

We now discuss the potential detectability of pulsars in Terzan 6 based on our searching results. The flux density of PSR~J1750--3116A can be estimated from the signal-to-noise ratio and the off-pulse noise \citep[see, e.g.,][]{2004hpa..book.....L}. The root-mean-square (RMS) variations in the
system noise
\begin{equation}
\Delta S_{\rm sys} =  \frac{T_{\rm sys}}{G\sqrt{n_{\rm p} t_{\rm bin} 
\Delta \nu }} = C \sigma_{\rm p},
\end{equation}
where $T_{\rm sys}=21.3$~K is the total system temperature, $G=2.0~{\rm K\,Jy^{-1}}$ is the telescope gain, $n_{\rm p}=2$ is the number of summed polarizations, $t_{\rm bin}$ is the
observation time per phase bin, $\Delta \nu$ is the effective bandwidth (2250~MHz after RFI zapping). As shown in this expression, these quantities are
related to the off-pulse RMS, $\sigma_{\rm p}$, by the scaling factor
$C$. From an uncalibrated pulse profile from which we measure $\sigma_{\rm p}$, 
this expression allows us to compute the scaling factor from which 
a mean flux density can be readily calculated by integrating the scaled
pulse profile. Using this approach, we find a mean flux density of 3.8~$\mu$Jy for the discovery observation (\autoref{fig:obs2}, where $t_{\rm bin}=223$~s) and 2.1~$\mu$Jy for the confirmation observation (\autoref{fig:obs1}, where $t_{\rm bin}=218$~s). {It should be noted that this approach is only valid when the off-pulse noise is Gaussian-like. However, as shown in the phase vs. time and phase vs. observing frequency panels in \autoref{fig:obs2} and \autoref{fig:obs1}, some RFI in both time and observing frequency domains remains present in the otherwise Gaussian-like off-pulse regions, indicating that the flux density is somewhat underestimated and should be treated as a lower limit.}

Pulsars typically have a negative spectral index \citep[around $-1.4$,][]{Bates+2013}, implying that their flux densities are higher at lower frequencies. Adopting an index of $-1.4$, we estimate the flux density of PSR~J1750--3116A at L-band $S_{\rm 1.4~{\rm GHz}}\sim 23~{\rm \mu Jy}$, suggesting that the pulsar may be bright enough to be detected with the L-band receivers of GBT (which has a 10-$\sigma$ sensitivity limit $\sim 13.2~{\rm \mu Jy}$ for a 4-hour integration at $1.44~{\rm GHz}$\footnote{Estimated by the GBT Sensitivity Calculator (\url{https://dss.gb.nrao.edu/calculator-ui/war/Calculator_ui.html}).}) or MeerKAT \citep[which has 10-$\sigma$ sensitivity limit $\sim 13~{\rm \mu Jy}$ for a 2.5-hour integration at $1.3~{\rm GHz}$,][]{Ridolfi+2021}. {Since observations at higher frequencies tend to detect pulsars with flatter spectra, the spectral index may be flatter than $-1.4$ for a typical pulsar. Therefore, the estimated flux density at L-band should be considered as an upper limit.} Even worse, significant scatter broadening and smearing due to the high dispersion measure might prohibit detection, especially for fastest MSPs  \citep[spin period $<{\rm 10~{\rm ms}}$;][]{Ridolfi+2021}. Additionally, since Terzan~6 is only $\sim 2\degree$ away from the Galactic center, {the sky temperatures at 1.4 GHz (L-band), 2.0 GHz (S-band) and 6 GHz (C-band) are 11.5 K, 6.4 K and 3.0 K for Terzan~6, respectively}\footnote{Calculated using \texttt{PyGDSM} \citep{PyGDSM}. The temperatures include a 2.725 K contribution from the cosmic microwave background.}. This extra contribution to the telescope's system temperature $T_{\rm sys}$ further complicates possible pulsar detection in L or S-band. These factors may explain why \cite{Lynch+2011} did not detect any pulsars in Terzan~6 during a 2-hour observation with the GBT {S-band (centered at $2~{\rm GHz}$)} receiver. Our C-band detection of PSR~J1750--3116A makes it possible for further follow-up L or S-band observations which make use of the now known DM. This may lead to new detections of pulsars in Terzan~6. 

To estimate how many MSPs might be detectable in the future, we can take the predicted number from \citet{Turk+2013} given in \autoref{tab:GCs} for Terzan 6, i.e. $\lambda=54$. Assuming a log-normal luminosity function \citep[see, e.g.,][]{2006ApJ...643..332F}, above some limiting luminosity, $L_{\rm min}$, the expected number of detectable pulsars  
\begin{equation}
\label{eq:ndet}
    N_{\rm det} \simeq \frac{\lambda}{2} \times {\rm erfc} \bigg[
    \frac{\log L_{\rm min}-\mu'}{\sqrt{2}\sigma'}
    \bigg],
\end{equation}
where $\mu'$ and $\sigma'$ are, respectively, the mean and standard deviation of the log-normal luminosity function. For a reference observing frequency of 1.4~GHz, we take the standard values of these parameters $\mu'=-1.1$ and $\sigma'=0.9$ that are consistent with GC populations \citep{2011MNRAS.418..477B}. The scaled 1.4~GHz flux density of 23~$\mu$Jy 
derived above for PSR~J1750--3116A corresponds to a luminosity of around 1.0~mJy~kpc$^2$. Setting this to be the value of $L_{\rm min}$ in \autoref{eq:ndet}, leads to an expected number of around 6 MSPs above
this limit. Further and deeper searches of Terzan~6 are indeed warranted.

\section{Conclusions}\label{sec:con}

We report the discovery of the $5.33~{\rm ms}$ pulsar~J1750--3116A in a targeted search of Terzan~6 with the GBT at C-band (4--8~GHz). This is the highest frequency discovery observation for a GC search to date. The pulsar has a dispersion measure of $\sim 383~{\rm pc~cm^{-3}}$, which is the second highest DM value among all GCs containing radio pulsars. The DM is compatible with that predicted by the Galactic free electron density model NE2001 \citep{Cordes+2002}, provided that Terzan~6 is $6.7~{\rm kpc}$ away. PSR~J1750--3116A is probably isolated, consistent with the Terzan~6's classification as a core-collapsed cluster, where dynamical interaction processes are more frequent and intense. These two factors lend support to the suggestion that the pulsar is associated with Terzan~6. The discovery of other pulsars with the same DM value of PSR J1750--3116A would provide definitive evidence of the association between PSR~J1750--3116A and Terzan~6.

We estimate the flux density of PSR~J1750--3116A to be $\sim 3~{\rm \mu Jy}$ at $6~{\rm GHz}$. Assuming a spectral index of $-1.4$, the flux density at $1.44~{\rm GHz}$ is inferred to be $23~{\rm \mu Jy}$, which is above the sensitivity limits for the GBT and MeerKAT L-band receivers. In consideration of the exceptionally high stellar encounter rate, more sensitive searches such as using GBT and MeerKAT \citep[e.g., TRAPUM, a MeerKAT large survey project,][]{Stappers+2016} are expected to result in further pulsar discoveries in Terzan~6. The more sensitive and long-term high-cadence follow-up observations can provide a phase-connected timing solution for PSR~J1750--3116A and hence localize its position to cross-check with the X-ray and $\gamma$-ray data.

\section*{Acknowledgements}
{We are grateful to the anonymous referee for their careful reading and insightful comments.}
S.J.G. would like to thank Evan Smith and Brenne Gregory for their assistance in transferring GBT archival data, Yang Chen and Yi-Wei Bao for providing the NVIDIA A100 GPU and De-Jiang Zhou for help with installing and using of the GPU-version \texttt{PRESTO}.
The National Radio Astronomy Observatory is a facility of the National Science Foundation operated under cooperative agreement by Associated Universities, Inc. The Green Bank Observatory is a facility of the National Science Foundation operated under cooperative agreement by Associated Universities, Inc.
S.J.G. acknowledges support from the National Natural Science Foundation of China (NSFC) under grant No.~123B2045.
X.D.L. acknowledges support from the National Key Research and Development Program of China (2021YFA0718500), the National Natural Science Foundation of China (NSFC) under grant No.~12041301, 12121003 and 12203051.
P.W. acknowledges support from the National Natural Science Foundation of China (NSFC) Programs No.~11988101, 12041303, the CAS Youth Interdisciplinary Team, the Youth Innovation Promotion Association CAS (id.~2021055), and the Cultivation Project for FAST Scientific Payoff and Research Achievement of CAMS-CAS.
P.Z. acknowledges support from the National Natural Science Foundation of China (NSFC) under grant No.~12273010.
L.Z. is supported by the National Natural Science Foundation of China  (NSFC) under grant No.~12103069.
Di Li is a New Cornerstone investigator.
The computation was made by using the facilities at the High-Performance Computing Center of Collaborative Innovation Center of Advanced Microstructures (Nanjing University).

\section*{Data Availability}
Raw PSRFITS format data can be accessed through \href{https://data.nrao.edu/portal/}{the NRAO Archive Interface} by contacting the GBT staff with project ID: AGBT19A\_477. {Ks-band image of Terzan~6 are based on data obtained from the ESO Science Archive Facility with DOI:\dataset[10.18727/archive/68]{\doi{10.18727/archive/68}}. Reduced data are available on the ScienceDB platform with DOI: \dataset[10.57760/sciencedb.12966]{\doi{10.57760/sciencedb.12966}}}.

\vspace{5mm}
\facilities{The Green Bank Observatory Robert C. Byrd Green Bank Telescope \citep{Prestage+2009}.}

\software{\texttt{PRESTO} \citep{Ransom+2011}, \texttt{TEMPO2} \citep{Nice+2015}, \texttt{DS9} \citep{DS9}, \texttt{PyGEDM} \citep{pygedm} and \texttt{PyGDSM} \citep{PyGDSM}.}

\bibliography{sample631}{}

\begin{thebibliography}{}
\expandafter\ifx\csname natexlab\endcsname\relax\def\natexlab#1{#1}\fi
\providecommand{\url}[1]{\href{#1}{#1}}
\providecommand{\dodoi}[1]{doi:~\href{http://doi.org/#1}{\nolinkurl{#1}}}
\providecommand{\doeprint}[1]{\href{http://ascl.net/#1}{\nolinkurl{http://ascl.net/#1}}}
\providecommand{\doarXiv}[1]{\href{https://arxiv.org/abs/#1}{\nolinkurl{https://arxiv.org/abs/#1}}}

\bibitem[{{Abdo} {et~al.}(2009){Abdo}, {Ackermann}, {Ajello}, {Atwood}, {Axelsson}, {Baldini}, {Ballet}, {Barbiellini}, {Baring}, {Bastieri}, {Baughman}, {Bechtol}, {Bellazzini}, {Berenji}, {Bignami}, {Blandford}, {Bloom}, {Bonamente}, {Borgland}, {Bregeon}, {Brez}, {Brigida}, {Bruel}, {Burnett}, {Caliandro}, {Cameron}, {Camilo}, {Caraveo}, {Carlson}, {Casandjian}, {Cecchi}, {{\c{C}}elik}, {Charles}, {Chekhtman}, {Cheung}, {Chiang}, {Ciprini}, {Claus}, {Cognard}, {Cohen-Tanugi}, {Cominsky}, {Conrad}, {Corbet}, {Cutini}, {Dermer}, {Desvignes}, {de Angelis}, {de Luca}, {de Palma}, {Digel}, {Dormody}, {do Couto e Silva}, {Drell}, {Dubois}, {Dumora}, {Edmonds}, {Farnier}, {Favuzzi}, {Fegan}, {Focke}, {Frailis}, {Freire}, {Fukazawa}, {Funk}, {Fusco}, {Gargano}, {Gasparrini}, {Gehrels}, {Germani}, {Giebels}, {Giglietto}, {Giordano}, {Glanzman}, {Godfrey}, {Grenier}, {Grondin}, {Grove}, {Guillemot}, {Guiriec}, {Hanabata}, {Harding}, {Hayashida}, {Hays}, {Hobbs}, {Hughes}, {J{\'o}hannesson}, {Johnson}, {Johnson},
  {Johnson}, {Johnson}, {Johnston}, {Kamae}, {Katagiri}, {Kataoka}, {Kawai}, {Kerr}, {Kn{\"o}dlseder}, {Kocian}, {Kramer}, {Kuss}, {Lande}, {Latronico}, {Lemoine-Goumard}, {Longo}, {Loparco}, {Lott}, {Lovellette}, {Lubrano}, {Madejski}, {Makeev}, {Manchester}, {Marelli}, {Mazziotta}, {McConville}, {McEnery}, {McLaughlin}, {Meurer}, {Michelson}, {Mitthumsiri}, {Mizuno}, {Moiseev}, {Monte}, {Monzani}, {Morselli}, {Moskalenko}, {Murgia}, {Nolan}, {Norris}, {Nuss}, {Ohsugi}, {Omodei}, {Orlando}, {Ormes}, {Paneque}, {Panetta}, {Parent}, {Pelassa}, {Pepe}, {Pesce-Rollins}, {Piron}, {Porter}, {Rain{\`o}}, {Rando}, {Ransom}, {Ray}, {Razzano}, {Rea}, {Reimer}, {Reimer}, {Reposeur}, {Ritz}, {Rochester}, {Rodriguez}, {Romani}, {Roth}, {Ryde}, {Sadrozinski}, {Sanchez}, {Sander}, {Saz Parkinson}, {Scargle}, {Schalk}, {Sgr{\`o}}, {Siskind}, {Smith}, {Smith}, {Spandre}, {Spinelli}, {Stappers}, {Starck}, {Striani}, {Strickman}, {Suson}, {Tajima}, {Takahashi}, {Tanaka}, {Thayer}, {Thayer}, {Theureau}, {Thompson}, {Thorsett},
  {Tibaldo}, {Torres}, {Tosti}, {Tramacere}, {Uchiyama}, {Usher}, {Van Etten}, {Vasileiou}, {Venter}, {Vilchez}, {Vitale}, {Waite}, {Wallace}, {Wang}, {Watters}, {Webb}, {Weltevrede}, {Winer}, {Wood}, {Ylinen}, \& {Ziegler}}]{Abdo+2009}
{Abdo}, A.~A., {Ackermann}, M., {Ajello}, M., {et~al.} 2009, Science, 325, 848, \dodoi{10.1126/science.1176113}

\bibitem[{{Bachetti} {et~al.}(2021){Bachetti}, {Pilia}, {Huppenkothen}, {Ransom}, {Curatti}, \& {Ridolfi}}]{Bachetti+2021}
{Bachetti}, M., {Pilia}, M., {Huppenkothen}, D., {et~al.} 2021, \apj, 909, 33, \dodoi{10.3847/1538-4357/abda4a}

\bibitem[{{Bagchi} {et~al.}(2011){Bagchi}, {Lorimer}, \& {Chennamangalam}}]{2011MNRAS.418..477B}
{Bagchi}, M., {Lorimer}, D.~R., \& {Chennamangalam}, J. 2011, \mnras, 418, 477, \dodoi{10.1111/j.1365-2966.2011.19498.x}

\bibitem[{{Bahramian} {et~al.}(2013){Bahramian}, {Heinke}, {Sivakoff}, \& {Gladstone}}]{Bahramian+2013}
{Bahramian}, A., {Heinke}, C.~O., {Sivakoff}, G.~R., \& {Gladstone}, J.~C. 2013, \apj, 766, 136, \dodoi{10.1088/0004-637X/766/2/136}

\bibitem[{{Barbuy} {et~al.}(1997){Barbuy}, {Ortolani}, \& {Bica}}]{Barbuy+1997}
{Barbuy}, B., {Ortolani}, S., \& {Bica}, E. 1997, \aaps, 122, 483, \dodoi{10.1051/aas:1997148}

\bibitem[{{Bates} {et~al.}(2013){Bates}, {Lorimer}, \& {Verbiest}}]{Bates+2013}
{Bates}, S.~D., {Lorimer}, D.~R., \& {Verbiest}, J.~P.~W. 2013, \mnras, 431, 1352, \dodoi{10.1093/mnras/stt257}

\bibitem[{{Boyles} {et~al.}(2011){Boyles}, {Lorimer}, {Turk}, {Mnatsakanov}, {Lynch}, {Ransom}, {Freire}, \& {Belczynski}}]{Boyles+2011}
{Boyles}, J., {Lorimer}, D.~R., {Turk}, P.~J., {et~al.} 2011, \apj, 742, 51, \dodoi{10.1088/0004-637X/742/1/51}

\bibitem[{{Cordes} \& {Lazio}(2002)}]{Cordes+2002}
{Cordes}, J.~M., \& {Lazio}, T.~J.~W. 2002, arXiv e-prints, astro, \dodoi{10.48550/arXiv.astro-ph/0207156}

\bibitem[{{Fahlman} {et~al.}(1995){Fahlman}, {Douglas}, \& {Thompson}}]{Fahlman+1995}
{Fahlman}, G.~G., {Douglas}, K.~A., \& {Thompson}, I.~B. 1995, \aj, 110, 2189, \dodoi{10.1086/117678}

\bibitem[{{Faucher-Gigu{\`e}re} \& {Kaspi}(2006)}]{2006ApJ...643..332F}
{Faucher-Gigu{\`e}re}, C.-A., \& {Kaspi}, V.~M. 2006, \apj, 643, 332, \dodoi{10.1086/501516}

\bibitem[{{Gratton} {et~al.}(2019){Gratton}, {Bragaglia}, {Carretta}, {D'Orazi}, {Lucatello}, \& {Sollima}}]{Gratton+2019}
{Gratton}, R., {Bragaglia}, A., {Carretta}, E., {et~al.} 2019, \aapr, 27, 8, \dodoi{10.1007/s00159-019-0119-3}

\bibitem[{{Harris}(1996)}]{Harris+1996}
{Harris}, W.~E. 1996, \aj, 112, 1487, \dodoi{10.1086/118116}

\bibitem[{{in't Zand} {et~al.}(2003){in't Zand}, {Hulleman}, {Markwardt}, {M{\'e}ndez}, {Kuulkers}, {Cornelisse}, {Heise}, {Strohmayer}, \& {Verbunt}}]{intZand+2003}
{in't Zand}, J.~J.~M., {Hulleman}, F., {Markwardt}, C.~B., {et~al.} 2003, \aap, 406, 233, \dodoi{10.1051/0004-6361:20030681}

\bibitem[{{Ivanova}(2013)}]{Ivanova+2013}
{Ivanova}, N. 2013, \memsai, 84, 123, \dodoi{10.48550/arXiv.1301.2203}

\bibitem[{{Joye} \& {Mandel}(2003)}]{DS9}
{Joye}, W.~A., \& {Mandel}, E. 2003, in Astronomical Society of the Pacific Conference Series, Vol. 295, Astronomical Data Analysis Software and Systems XII, ed. H.~E. {Payne}, R.~I. {Jedrzejewski}, \& R.~N. {Hook}, 489

\bibitem[{{Kremer} {et~al.}(2023){Kremer}, {Fuller}, {Piro}, \& {Ransom}}]{Kremer+2023}
{Kremer}, K., {Fuller}, J., {Piro}, A.~L., \& {Ransom}, S.~M. 2023, \mnras, 525, L22, \dodoi{10.1093/mnrasl/slad088}

\bibitem[{{Kremer} {et~al.}(2021){Kremer}, {Piro}, \& {Li}}]{Kremer+2021}
{Kremer}, K., {Piro}, A.~L., \& {Li}, D. 2021, \apjl, 917, L11, \dodoi{10.3847/2041-8213/ac13a0}

\bibitem[{{Kremer} {et~al.}(2022){Kremer}, {Ye}, {K{\i}ro{\u{g}}lu}, {Lombardi}, {Ransom}, \& {Rasio}}]{Kremer+2022}
{Kremer}, K., {Ye}, C.~S., {K{\i}ro{\u{g}}lu}, F., {et~al.} 2022, \apjl, 934, L1, \dodoi{10.3847/2041-8213/ac7ec4}

\bibitem[{{Leahy} {et~al.}(1983){Leahy}, {Darbro}, {Elsner}, {Weisskopf}, {Sutherland}, {Kahn}, \& {Grindlay}}]{Leahy+1983}
{Leahy}, D.~A., {Darbro}, W., {Elsner}, R.~F., {et~al.} 1983, \apj, 266, 160, \dodoi{10.1086/160766}

\bibitem[{{Lorimer} \& {Kramer}(2004)}]{2004hpa..book.....L}
{Lorimer}, D.~R., \& {Kramer}, M. 2004, {Handbook of Pulsar Astronomy}, Vol.~4

\bibitem[{{Lynch} {et~al.}(2011){Lynch}, {Ransom}, {Freire}, \& {Stairs}}]{Lynch+2011}
{Lynch}, R.~S., {Ransom}, S.~M., {Freire}, P. C.~C., \& {Stairs}, I.~H. 2011, \apj, 734, 89, \dodoi{10.1088/0004-637X/734/2/89}

\bibitem[{{Lyne} {et~al.}(1996){Lyne}, {Manchester}, \& {D'Amico}}]{1996ApJ...460L..41L}
{Lyne}, A.~G., {Manchester}, R.~N., \& {D'Amico}, N. 1996, \apjl, 460, L41, \dodoi{10.1086/309972}

\bibitem[{{McCarver} {et~al.}(2024){McCarver}, {Maccarone}, {Ransom}, {Clarke}, {Giacintucci}, {Peters}, {Polisensky}, {Nyland}, {Gautam}, {Freire}, \& {Rangelov}}]{McCarver+2024}
{McCarver}, A.~V., {Maccarone}, T.~J., {Ransom}, S.~M., {et~al.} 2024, \apj, 969, 30, \dodoi{10.3847/1538-4357/ad4461}

\bibitem[{{Minniti}(2018)}]{Minniti+VVVX}
{Minniti}, D. 2018, in Astrophysics and Space Science Proceedings, Vol.~51, The Vatican Observatory, Castel Gandolfo: 80th Anniversary Celebration, ed. G.~{Gionti} \& J.-B. {Kikwaya Eluo}, 63, \dodoi{10.1007/978-3-319-67205-2_4}

\bibitem[{{Nice} {et~al.}(2015){Nice}, {Demorest}, {Stairs}, {Manchester}, {Taylor}, {Peters}, {Weisberg}, {Irwin}, {Wex}, \& {Huang}}]{Nice+2015}
{Nice}, D., {Demorest}, P., {Stairs}, I., {et~al.} 2015, {Tempo: Pulsar timing data analysis}, Astrophysics Source Code Library, record ascl:1509.002

\bibitem[{{Padmanabh} {et~al.}(2024){Padmanabh}, {Ransom}, {Freire}, {Ridolfi}, {Taylor}, {Choza}, {Clark}, {Abbate}, {Bailes}, {Barr}, {Buchner}, {Burgay}, {DeCesar}, {Chen}, {Corongiu}, {Champion}, {Dutta}, {Geyer}, {Hessels}, {Kramer}, {Possenti}, {Stairs}, {Stappers}, {Venkatraman Krishnan}, {Vleeschower}, \& {Zhang}}]{Padmanabh+2024}
{Padmanabh}, P.~V., {Ransom}, S.~M., {Freire}, P.~C.~C., {et~al.} 2024, \aap, 686, A166, \dodoi{10.1051/0004-6361/202449303}

\bibitem[{{Pooley} {et~al.}(2003){Pooley}, {Lewin}, {Anderson}, {Baumgardt}, {Filippenko}, {Gaensler}, {Homer}, {Hut}, {Kaspi}, {Makino}, {Margon}, {McMillan}, {Portegies Zwart}, {van der Klis}, \& {Verbunt}}]{Pooley+2003}
{Pooley}, D., {Lewin}, W. H.~G., {Anderson}, S.~F., {et~al.} 2003, \apjl, 591, L131, \dodoi{10.1086/377074}

\bibitem[{{Predehl} {et~al.}(1991){Predehl}, {Hasinger}, \& {Verbunt}}]{Predehl+1991}
{Predehl}, P., {Hasinger}, G., \& {Verbunt}, F. 1991, \aap, 246, L21

\bibitem[{{Prestage} {et~al.}(2009){Prestage}, {Constantikes}, {Hunter}, {King}, {Lacasse}, {Lockman}, \& {Norrod}}]{Prestage+2009}
{Prestage}, R.~M., {Constantikes}, K.~T., {Hunter}, T.~R., {et~al.} 2009, IEEE Proceedings, 97, 1382, \dodoi{10.1109/JPROC.2009.2015467}

\bibitem[{{Price}(2016)}]{PyGDSM}
{Price}, D.~C. 2016, {PyGDSM: Python interface to Global Diffuse Sky Models}, Astrophysics Source Code Library, record ascl:1603.013

\bibitem[{{Price} {et~al.}(2021){Price}, {Flynn}, \& {Deller}}]{pygedm}
{Price}, D.~C., {Flynn}, C., \& {Deller}, A. 2021, \pasa, 38, e038, \dodoi{10.1017/pasa.2021.33}

\bibitem[{{Ransom}(2011)}]{Ransom+2011}
{Ransom}, S. 2011, {PRESTO: PulsaR Exploration and Search TOolkit}, Astrophysics Source Code Library, record ascl:1107.017

\bibitem[{{Ransom} {et~al.}(2002){Ransom}, {Eikenberry}, \& {Middleditch}}]{Ransom+2002}
{Ransom}, S.~M., {Eikenberry}, S.~S., \& {Middleditch}, J. 2002, \aj, 124, 1788, \dodoi{10.1086/342285}

\bibitem[{{Ridolfi} {et~al.}(2021){Ridolfi}, {Gautam}, {Freire}, {Ransom}, {Buchner}, {Possenti}, {Venkatraman Krishnan}, {Bailes}, {Kramer}, {Stappers}, {Abbate}, {Barr}, {Burgay}, {Camilo}, {Corongiu}, {Jameson}, {Padmanabh}, {Vleeschower}, {Champion}, {Chen}, {Geyer}, {Karastergiou}, {Karuppusamy}, {Parthasarathy}, {Reardon}, {Serylak}, {Shannon}, \& {Spiewak}}]{Ridolfi+2021}
{Ridolfi}, A., {Gautam}, T., {Freire}, P.~C.~C., {et~al.} 2021, \mnras, 504, 1407, \dodoi{10.1093/mnras/stab790}

\bibitem[{{Stappers} \& {Kramer}(2016)}]{Stappers+2016}
{Stappers}, B., \& {Kramer}, M. 2016, in MeerKAT Science: On the Pathway to the SKA, 9, \dodoi{10.22323/1.277.0009}

\bibitem[{{Trager} {et~al.}(1995){Trager}, {King}, \& {Djorgovski}}]{Trager+1995}
{Trager}, S.~C., {King}, I.~R., \& {Djorgovski}, S. 1995, \aj, 109, 218, \dodoi{10.1086/117268}

\bibitem[{{Turk} \& {Lorimer}(2013)}]{Turk+2013}
{Turk}, P.~J., \& {Lorimer}, D.~R. 2013, \mnras, 436, 3720, \dodoi{10.1093/mnras/stt1850}

\bibitem[{{Valenti} {et~al.}(2007){Valenti}, {Ferraro}, \& {Origlia}}]{Valenti+2007}
{Valenti}, E., {Ferraro}, F.~R., \& {Origlia}, L. 2007, \aj, 133, 1287, \dodoi{10.1086/511271}

\bibitem[{{van den Berg} {et~al.}(2024){van den Berg}, {Homan}, {Heinke}, {Pooley}, {Wijnands}, {Bahramian}, \& {Miller-Jones}}]{vandenBerg+2024}
{van den Berg}, M., {Homan}, J., {Heinke}, C.~O., {et~al.} 2024, \apj, 966, 217, \dodoi{10.3847/1538-4357/ad2f3d}

\bibitem[{{Verbunt} \& {Freire}(2014)}]{Verbunt+2014}
{Verbunt}, F., \& {Freire}, P. C.~C. 2014, \aap, 561, A11, \dodoi{10.1051/0004-6361/201321177}

\bibitem[{{Verbunt} \& {Hut}(1987)}]{Verbunt+1987}
{Verbunt}, F., \& {Hut}, P. 1987, in The Origin and Evolution of Neutron Stars, ed. D.~J. {Helfand} \& J.~H. {Huang}, Vol. 125, 187

\bibitem[{{Vleeschower} {et~al.}(2024){Vleeschower}, {Corongiu}, {Stappers}, {Freire}, {Ridolfi}, {Abbate}, {Ransom}, {Possenti}, {Padmanabh}, {Balakrishnan}, {Kramer}, {Venkatraman Krishnan}, {Zhang}, {Bailes}, {Barr}, {Buchner}, \& {Chen}}]{Vleeschower+2024}
{Vleeschower}, L., {Corongiu}, A., {Stappers}, B.~W., {et~al.} 2024, \mnras, 530, 1436, \dodoi{10.1093/mnras/stae816}

\bibitem[{{Wu} {et~al.}(2023){Wu}, {Pan}, {Qian}, {Ransom}, {Wang}, {Yan}, {Luo}, {Zhang}, {Li}, {Yin}, {Li}, {Li}, {Dai}, {Li}, {Zhang}, {Liu}, \& {Pan}}]{Wu+2023}
{Wu}, Y., {Pan}, Z., {Qian}, L., {et~al.} 2023, arXiv e-prints, arXiv:2312.06067, \dodoi{10.48550/arXiv.2312.06067}

\bibitem[{{Yao} {et~al.}(2017){Yao}, {Manchester}, \& {Wang}}]{Yao+2017}
{Yao}, J.~M., {Manchester}, R.~N., \& {Wang}, N. 2017, \apj, 835, 29, \dodoi{10.3847/1538-4357/835/1/29}

\bibitem[{{Ye} {et~al.}(2024){Ye}, {Kremer}, {Ransom}, \& {Rasio}}]{Ye+2024}
{Ye}, C.~S., {Kremer}, K., {Ransom}, S.~M., \& {Rasio}, F.~A. 2024, \apj, 961, 98, \dodoi{10.3847/1538-4357/ad089a}

\bibitem[{{Zhou} {et~al.}(2024){Zhou}, {Wang}, {Li}, {Fang}, {Miao}, {Freire}, {Zhang}, {Zhang}, {Chen}, {Feng}, {Xiao}, {Xie}, {Zhang}, {Jin}, {Wang}, {Ke}, {Guo}, {Zhao}, {Niu}, {Zhu}, {Xue}, {Wang}, {Wu}, {Gan}, {Sun}, {Wang}, {Zhang}, {Zhang}, {Cao}, \& {Lu}}]{Zhou+2023}
{Zhou}, D., {Wang}, P., {Li}, D., {et~al.} 2024, Science China Physics, Mechanics, and Astronomy, 67, 269512, \dodoi{10.1007/s11433-023-2362-x}

\end{thebibliography}
\bibliographystyle{aasjournal}



\end{document}